\documentclass[letterpaper, 10 pt, conference]{ieeeconf}
\usepackage{cite} 
\usepackage{multirow}
\usepackage[utf8]{inputenc}
\usepackage[misc]{ifsym}
\usepackage{graphicx}
\usepackage{tikz}
\usepackage{algorithm}
\usepackage{algorithmic}
\usepackage{pifont}
\usepackage[misc]{ifsym}
\usepackage{tikz}
\usepackage{pdfpages}
\usepackage{graphicx}
\usepackage{amsmath}
\usepackage[utf8]{inputenc}
\usetikzlibrary{arrows,shapes,positioning,shadows,trees,calc,decorations.markings}

\usepackage{amssymb}
\usepackage{hyperref}

\makeatletter
\newcommand{\printfnsymbol}[1]{%
  \textsuperscript{\@fnsymbol{#1}}%
}
\makeatother

\IEEEoverridecommandlockouts                

\title{\LARGE \bf
Glioblastoma Multiforme Prognosis: MRI Missing Modality Generation, Segmentation and Radiogenomic Survival Prediction
}

\author{Mobarakol Islam, Navodini Wijethilake  and Hongliang Ren
\thanks{M. Islam is with NUS Graduate School for Integrative Sciences and Engineering (NGS), National University of Singapore, Singapore}

\thanks{M. Islam and H. Ren are with Dept. of Biomedical Engineering, National University of Singapore, Singapore; Corresponding author: Hongliang Ren, hlren@ieee.org http://labren.org}
\thanks{N. Wijethilake  is with dept. of Electronics and Telecommunications, University of Moratuwa, Srilanka}
}

\usepackage{amsmath}
\begin{document}

\maketitle
\thispagestyle{empty}
\pagestyle{empty}

\begin{abstract}
The accurate prognosis of Glioblastoma Multiforme (GBM) plays an essential role in planning correlated surgeries and treatments. The conventional models of survival prediction rely on radiomic features using magnetic resonance imaging (MRI). In this paper, we propose a radiogenomic overall survival (OS) prediction approach by incorporating gene expression data with radiomic features such as shape, geometry, and clinical information.  We exploit TCGA (The Cancer Genomic Atlas) dataset and synthesize the missing MRI modalities using a fully convolutional network (FCN) in a conditional Generative Adversarial Network (cGAN). Meanwhile, the same FCN architecture enables the tumor segmentation from the available and the synthesized MRI modalities. The proposed FCN architecture comprises octave convolution (OctConv) and a novel decoder, with skip connections in spatial and channel squeeze \& excitation (skip-scSE) block. The OctConv can process low and high-frequency features individually and improve model efficiency by reducing channel-wise redundancy. Skip-scSE applies spatial and channel-wise excitation to signify the essential features and reduces the sparsity in deeper layers learning parameters using skip connections. The proposed approaches are evaluated by comparative experiments with state-of-the-art models in synthesis, segmentation, and overall survival (OS) prediction. We observe that adding missing MRI modality improves the segmentation prediction, and expression levels of gene markers have a high contribution in the GBM prognosis prediction, and fused radiogenomic features boost the OS estimation.    
\end{abstract}

\maketitle

\section{Introduction}
Glioblastoma Multiforme (GBM), as the lethal primary brain tumor, has one of the worst survival rates out of all the human cancers. There are 26.5\% of patients diagnosed with GBM with an average survival rate of 2-year and median survival of 14-16 months with radiotherapy and temozolomide \cite{stupp2005radiotherapy, bleeker2012recent}. In current clinical practise, clinicians tend to estimate the survival, with their experience, by analysing the images and considering other clinical factors such as age, gender. However, researches show that these decisions and estimations are too much biased and optimistic \cite{moghtadaei2014predicting}. Therefore, early diagnosis and accurate prognostication of GBM progression demand in-depth research utilizing modern data driven technologies. 

In conventional research procedures, radiological data such as magnetic resonance (MR) images are used for delineating tumor regions and extracting quantitative imaging features to estimate the survival rate, known as radiomics \cite{islam2018glioma,lao2017deep,prasanna2017radiomic}. However, recent studies \cite{gonzalez2013personalized, weller2010mgmt} find associations of gene biomarkers with the prognosis of glioma and treatment planning, which comes under genomics. Thus, current studies focus on the associations between radiomics and underlying genomics, which is known as radiogenomics. The correlations between gene biomarkers such as MGMT methylation, IDH mutations, gene microarray data, and radiomics are analyzed under radiogenomics \cite{lee2015associating,xi2018radiomics,korfiatis2016mri,incoronato2017radiogenomic,wijethilake2020radiogenomics,wijethilake2020mbec}. Nevertheless, the MGMT methylation status of GBM patients is incorporated with radiomics to analyze glioma patients' survival \cite{tixier2019preoperative}. Further, Chaddad et al. \cite{chaddad2018novel} incorporate the expression of gene markers with radiomic for survival prediction. Correspondingly, in this work, we incorporate gene expression profiles with radiomic features to produce a more predictive outcome for survival estimation.   

Magnetic resonance imaging (MRI) is frequently utilized to follow-up brain tumors such as glioma, glioblastoma as a powerful noninvasive diagnostic tool. Based on different contrasts and functional features, there are several MRI modalities such as T1 or T1-contrast (spin-lattice relaxation), T2 (spin-spin relaxation), and Flair (T2 fluid-attenuated inversion recovery) \cite{tseng2017joint}. Different tissue structures and underlying anatomy are visible in different modalities. For an example, grey and white matter are visible in T1-weighted brain scan, T2 weighted brain scan locates the lesion by imaging fluids in the cortical tissue while Flair images locate the lesions through water suppression \cite{ma2018tailored}. However, the full spectrum of multi-modality MRI scans is unavailable in most cases due to the scanning cost, time, scanner availability, and patient comfort. A subset may be corruptive by excessive noise or artifacts.
Moreover, remarkable longitudinal studies like ADNI \cite{mueller2005alzheimer} requires scanning changes over time, which may exist in different acquisition protocol. In this case, remarkable modalities could be missing or corrupted in the large cohort study due to the improvement of MRI scanners or changing the observation from one tissue to another. This information missing can cause restraints in MRI analysis, diagnosis, and research studies. Thus, synthesizing the missing or corrupted modalities is essential in image related diagnosis, therapeutic planning, and research tasks such as segmentation, detection, multi-modal registration, and classification.

In this study, we use TCGA\footnote[1]{https://www.cancer.gov/tcga} (the genomic cancer atlas) dataset and design a radiogenomic model to
estimate survival duration. In the cohort, there are very few cases with all four modalities. We synthesize the missing modalities, segment the tumor regions using existing and synthesized modalities, and estimate the overall survival (OS) using the radiogenomic model. We exploit BraTS-2017 \cite{menze2015multimodal, bakas2017advancing, bakas2017segmentation} dataset to train and validate the synthesize and segmentation model.

\subsection{Related Works}
Recently, a deep convolutional neural network (DCNN) shows state-of-the-art (SOTA) performance in the biomedical applications of detection, classification, segmentation, and synthesizing. In particular, the generative adversarial network (GAN) performs outstanding performance in the task of an image to image translation.   
\subsubsection{Image Synthesis}
There are several studies on designing an image synthesize model using GAN from computer vision to medical imaging applications. CVAE-GAN \cite{bao2017cvae}, a conditional variational autoencoder GAN  combines a variational auto-encoder with a GAN for synthesizing images. BicycleGAN \cite{zhu2017toward} adopts CVAE-GAN \cite{bao2017cvae} and conditional latent regressor GAN (CLR-GAN) \cite{donahue2016adversarial, dumoulin2016adversarially} for multi-modal images to single image mapping. Conditional adversarial network (cGAN) \cite{isola2017image} learns the nonlinear mapping of the input to output image and trains a loss function to apply the same generic approach to different problems synthesizing, reconstructing, and colorizing images. Dar et al \cite{dar2018image} adopt cGAN \cite{isola2017image} to translate MRI T2 weighted from source T1-weighted images. However, these studies exploit a simple, fully convolutional network (FCN) or vanilla UNet \cite{ronneberger2015u} as a generator. Moreover, most of these works exploit shallow encoder type architecture or patch discriminator instead of full encoder-decoder.

\subsubsection{Tumor Segmentation and Prognosis}
In the past years, brain tumor segmentation and overall survival prediction models have been associative remarkably. An ensemble 3D UNet and linear regression models can segment and predict survival \cite{feng2018brain}. Similarly, tumor progression is estimated using a VGG-16 based FCN and random forest models \cite{puybareau2018segmentation} and multi-modal
PixelNet and artificial neural network (ANN) \cite{islam2018glioma}. However, the prediction performance of these models is not satisfactory, as only radiomic features are used to train the model. Apart from using radiomics, survival prediction of GBM patients by using multidimensional genomic data is used \cite{kim2013identification}. Recently, the use of radiogenomics, integrating imaging with gene expression profiles proliferated. A radiogenomic model design to predict risk estimation using logistic regression model \cite{qian2018radiogenomics}. However, the study was not estimating the exact survival days.

\begin{figure}
\centering
\includegraphics[width=0.5\textwidth]{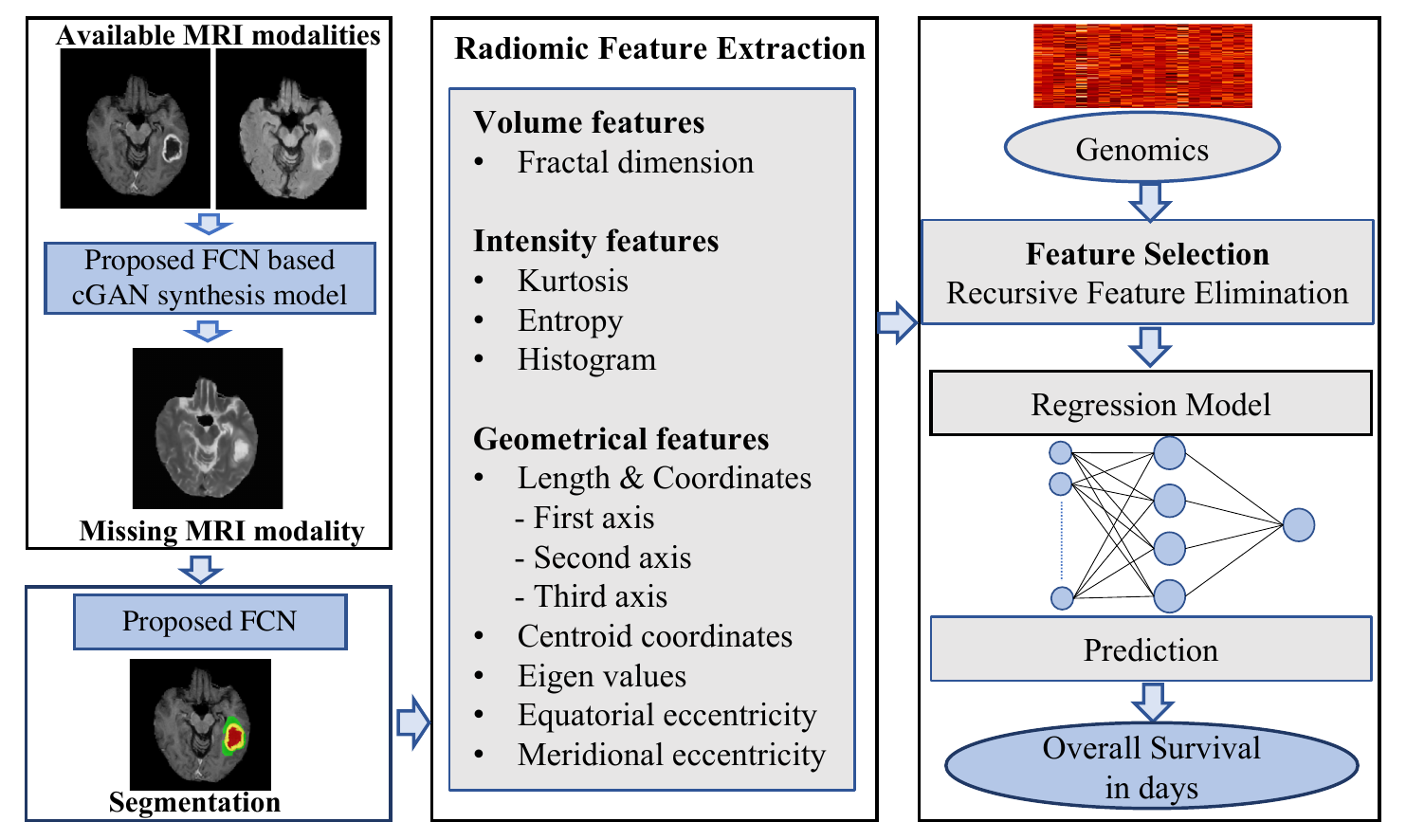}
\caption{The workflow of synthesizing, segmentation and radiogenomic analysis in this study.}
\label{fig:workflow}
\end{figure}
\subsubsection{FCN Model}
Fully convolutional networks (FCNs) show the state-of-the-art performance of visual tasks by capturing the channel's hierarchical features and spatial information with receptive fields. "Squeeze-and-Excitation" (SE) network \cite{hu2018squeeze} performs channel-wise feature recalibration to emphasize informative features and suppress weak ones. Competitive SE architecture \cite{hu2018competitive} re-weights the feature maps with channel-wise attention for residual connection. Spatial and channel squeeze \& excitation (scSE) \cite{roy2018concurrent} in FCN applies excitation in spatial and channel-wise jointly. However, we hypothesize that squeezing weak features towards zero may increase the sparsity in deeper layers prone to poor learning \cite{uhrig2017sparsity}. Most recently, Octave convolution (OctConv) \cite{chen2019drop} introduces a separate convolution of high and low-frequency features encoded into an image that reduces the channel-wise redundancy. Low-frequency feature maps are upsampled to high-frequency maps and concatenated in the last layer to predict the output. 
% To eliminate the potential undesired effects of zero paddings, Partial Convolution based Padding \cite{DBLP:journals/corr/abs-1811-11718} re-weights the convolution output with the fraction of the padded region by convolution sliding window.

\begin{figure*}
\centering
\includegraphics[width=1\textwidth
]{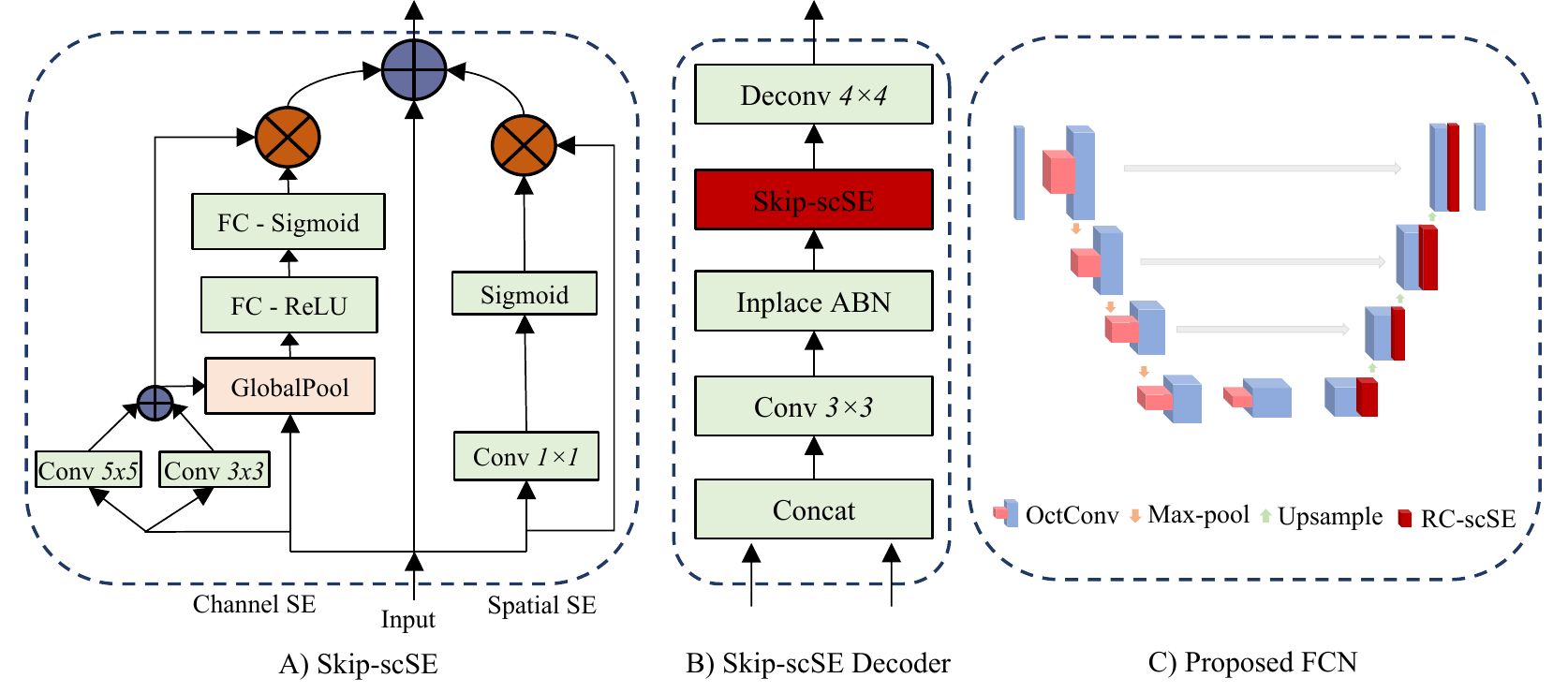}
\caption{Proposed modules and model architecture.{(A)} Skip connection with scSE (skip-scSE) module. {(B)} Proposed decoder block. Skip-scSE module after a convolution followed by inplace ABN and a deconvolution layer at the end of each block. {(C)} Our FCN architecture with encode-decoder block of OctConv.}
\label{fig:archi_octave}
\end{figure*}

\subsection{Contributions}
In this study, we propose an FCN by leveraging OctConv and a novel decoder with a skip SE unit of spatial and channel-wise recalibration. Proposed FCN is utilized to generate missing MRI modalities and segment the tumor for TCGA using BraTS-2017 \cite{menze2015multimodal, bakas2017advancing, bakas2017segmentation} dataset. The workflow of the study is illustrated in Figure \ref{fig:workflow}. Our contributions are summarized as follows:
\begin{itemize}
\item[--] Propose an FCN architecture using octave partial convolution and skip connection with scSE (skip-scSE).
\item[--] Design a novel decoder efficiently using a skip-scSE module.
\item[--] Design a multi-modal MRI to single MRI modality synthesizing model using proposed F
CN as a generator and encoder-decoder discriminator using the cGAN learning approach.
\item[--] Generate missing modalities for the TCGA dataset using the available modality (or modalities) and predict segmentation using both available and synthesized modalities.
\item[--] Design a Radiogenomic model by combining radiomic features such as intensity, shape, and geometry with gene expression profiling data.
\end{itemize}
\section{Methods}
We propose a fully convolutional network (FCN) model with Octave convolution \cite{chen2019drop} and a novel decoder based on channel and spatial excitation. The model applies for both synthesizing MRI missing modality and tumor segmentation tasks, as shown in Figure \ref{fig:syn_seg}. The radiomic features are extracted from the segmented tumor and combined with genomic data to predict survival rate using an artificial neural network (ANN) and support vector machine (SVM) model.  

\subsection{Proposed FCN}
\begin{figure}
\centering
\includegraphics[width=0.5\textwidth]{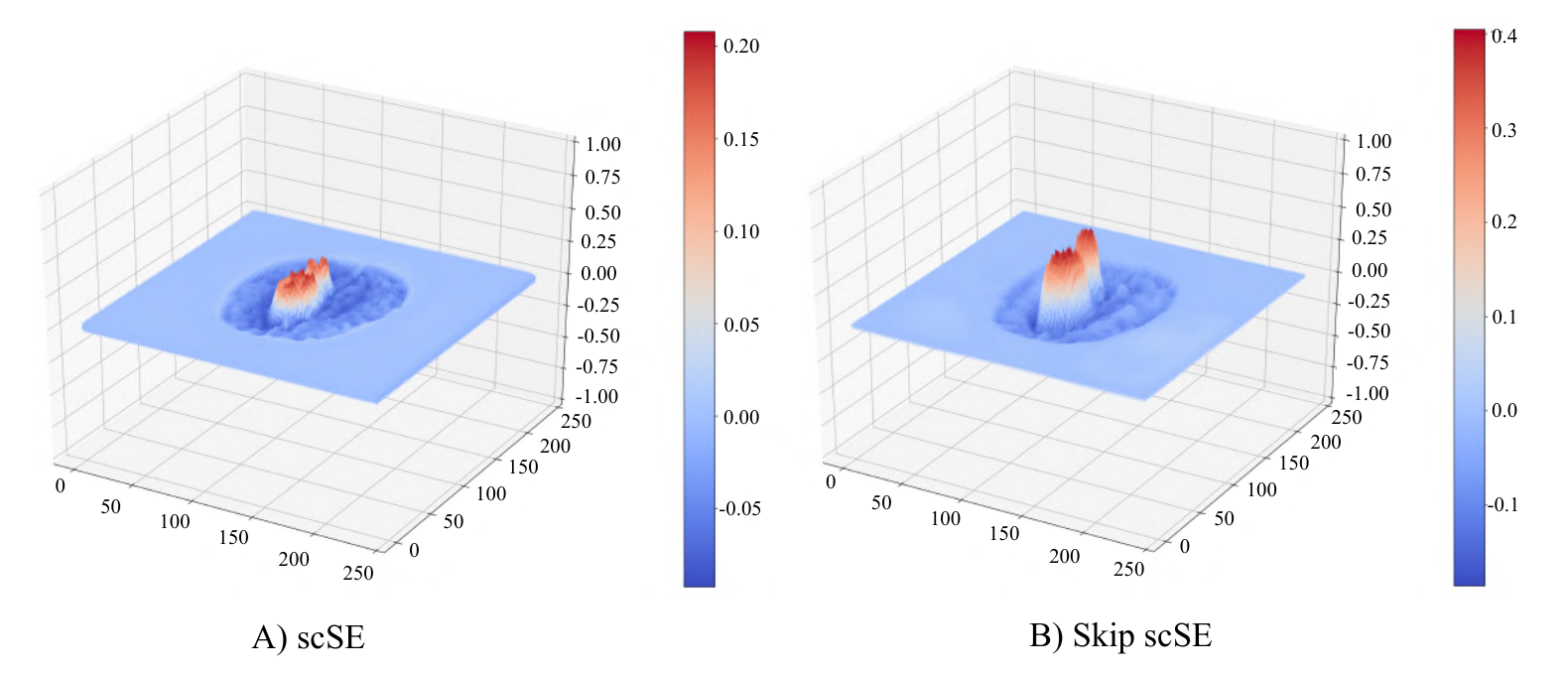}
\caption{Similar feature maps with original (A) scSE and the proposed (B) skip scSE. Skip scSE excite the feature maps higher range than original scSE. The feature maps are visualized from the trained models of scSE and skip-scSE.}
\label{fig:skip_scse}
\end{figure}

\subsubsection{Skip-scSE}
Spatial and channel "Squeeze \& Excitation" (scSE) \cite{roy2018concurrent, hu2018squeeze} recalibrates the feature maps to suppress the weak features and signify the essential features. Competitive inner-imaging SE \cite{hu2018competitive} enhances the learning and re-weight the channel of residual mapping to alleviate the redundancy in feature maps. Selective kernel convolution \cite{DBLP:journals/corr/abs-1903-06586} uses convolutions of different kernel sizes to enable neurons to adjust their receptive fields (RF) sizes adaptively. However, recalibrating non-significant features towards zero with scSE may increase the sparsity in deeper layers as well as reduce the parameter learning \cite{uhrig2017sparsity}. We observe that
adding skip connection with scSE retains weak features and boosts the significant features' excitation. We design the skip-scSE module by adding a skip connection and inner-imaging layers with scSE, as in Figure \ref{fig:archi_octave} (A). Inner-imaging improves the feature representation before applying channel excitation. Figure \ref{fig:skip_scse} shows that our skip-scSE boosts the excitation of the feature maps to [-0.1 to 0.4] where the range of feature with original scSE is [-0.05 to 0.20].

\subsubsection{Network Architecture}
Our FCN model consists of encoder blocks similar to UNet \cite{ronneberger2015u} with octave convolution and a novel decoder, including the skip-scSE module. Each encoder block consists of two 3x3  OctConv, followed by ReLU and max-pooling of stride 2. OctConv capable of processing higher and lower frequency information separately. For example, the high-frequency contextual features like fine details and low-frequency features like global structure, are encoded into an image. It applies group convolution to reduce channel-wise redundancy. 

Skip-scSE decoder concatenates the encoder output and skip connection followed by sequential layers of 3x3 convolution, inplace activated batch-norm (ABN) \cite{rota2018place}, skip-scSE module and 4x4 deconvolution. Figure \ref{fig:archi_octave} (B) and \ref{fig:archi_octave} (C) represent the skip-scSE decoder and proposed FCN model.

\subsection{MRI-MRI translation}
\label{sec:mri_to_mri}
\begin{figure*}
\centering
\includegraphics[width=1.0\textwidth]{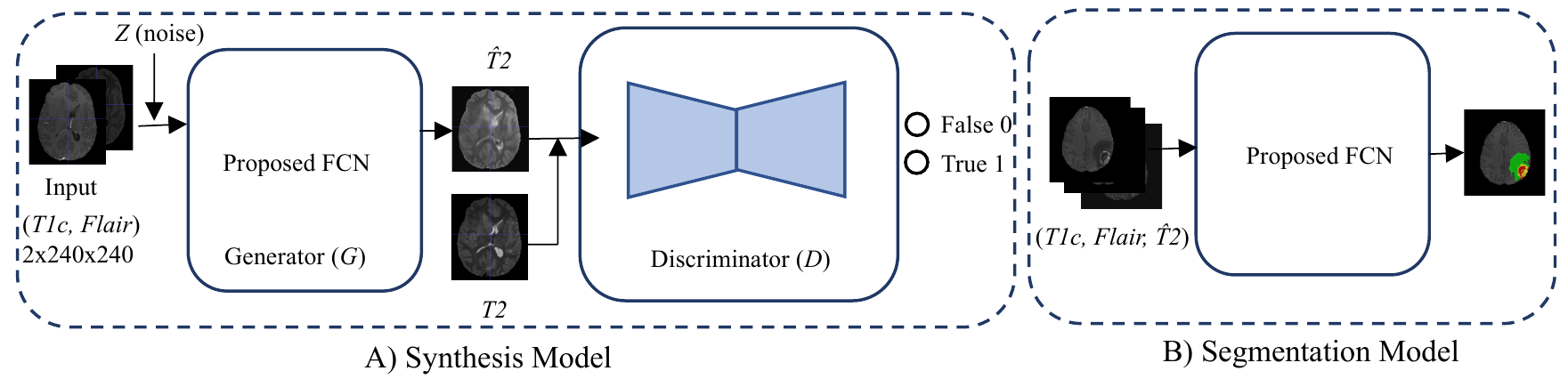}
\caption{Our approaches for MRI translation and segmentation. (A) Proposed FCN uses as the generator network in cGAN; (B) Our model to segment the tumor from multi-modal MRI. Red, yellow and green color in the predictio
n represent the tumor regions of necrotic, enhancing, and edema.}
\label{fig:syn_seg}
\end{figure*}

Generative Adversarial Network (GAN) contains a Generator ($G$) and a Discriminator ($D$), two sub-network where a generator network learns to map the provided input sequence and a discriminator network distinguishes between feature maps between a target distribution and the output of generator network. We adopt proposed FCN as $G$ network in conditional GAN \cite{isola2017image} and encoder-decoder light-weight design of $D$ architecture as \cite{islam2019real} (see Figure \ref{fig:syn_seg}A).

In the training time, latent variable (random noise) $Z$ with input modalities such as $T1ce$ and $Flair$ are fed to the $G$ network which aims to translate identical modality of $\hat{T2}$ from the real contrast of $T2$. $D$ network evolves to distinguish between generated modality and real modality images. The prediction of $D$ is 1 or 0 where 1 denotes real modality and 0 denotes translated modality. Both networks $G$ and $D$ learn simultaneously in training phase by minimizing and maximizing adversarial loss respectively. We follow same objective function as pix2pix \cite{isola2017image} to optimize our model. The objective function can be formulated as- 
\begin{eqnarray}\label{eq:overall_obj}
G_{obj} =\text{arg} \underset{G}{ \text{ min}} \underset{D}{ \text{ max }} \mathcal{L}_\text{$cGAN$}(G,D) + \lambda\mathcal{L}_\text{$L1$}(G)
\end{eqnarray}

where $\mathcal{L}_\text{$cGAN$}(G,D)$ is the objective function of the conditional GAN. The $G$ model, the generator, tries to minimize the objective function against the adversarial model $D$ that tries to maximize it. $\mathcal{L}_\text{$L1$}(G)$ is the distance loss, and \(\lambda\) is the adversarial factor to tune the distance loss.

\subsection{Survival Prediction}
Survival prediction is performed with genomics, radiomics, and radiogenomics. Radiomic features are extracted from the segmented tumor and fused with gene expression profiling data to estimate the OS rate.

\subsubsection{Feature Extraction}
\begin{figure}[!h]
\centering
\includegraphics[width=0.5\textwidth]{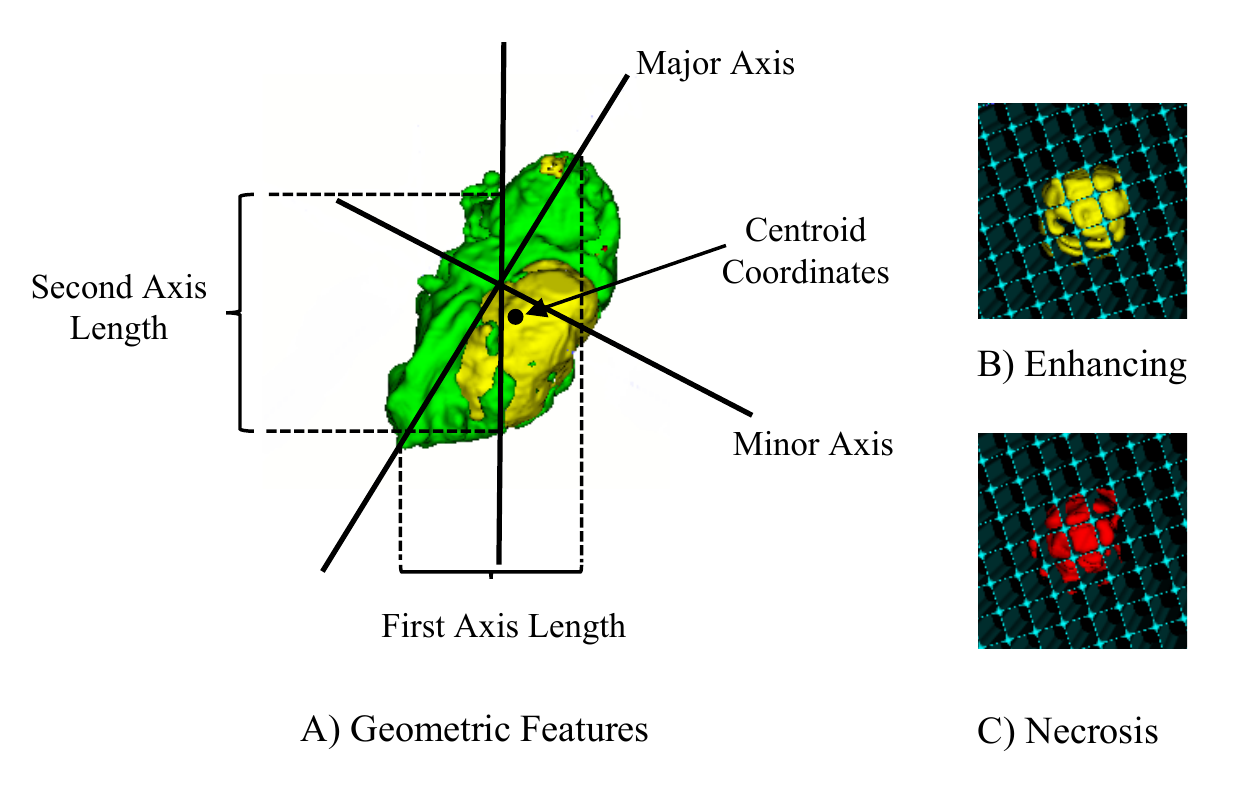}
\caption{ {(A)} Geometrical features extracted from the segmentation;{(B)} Fractal analysis for Enhance tumor; {(C)} Fractal analysis for necrosis. The Box-counting method \cite{theiler1990estimating} is used to determine the fractal properties of the 3D segmented MRI.}
\label{fig:boxcount}
\end{figure}

%Geometric information plays a vital role in the survival of Glioblastoma patients. Patients with temporal or parietal lobe tumors have poor prognosis than patients with frontal lobe tumors \cite{simpson1993influence}, as shown in figure \ref{fig:Survival}.
For predicting survival rate, geometry, shape, and intensity features are extracted from segmented tumor and corresponding sub-regions inside MRI. The geometrical features we extracted are coordinates and lengths of the first axis, second axis and third axis, centroid coordinates, eigenvalues,  meridional and equatorial eccentricity for Necrosis, Tumor core and whole tumor sub-regions of the tumor by following these works\cite{islam2017fully, farag2011evaluation, ko2013tree}. Meridional eccentricity is the eccentricity measured at the section containing both the longest and the shortest axes. Equatorial eccentricity is measured at the section through the center, perpendicular to the polar axis. Figure \ref{fig:boxcount} (A) shows the measurement of the extracted geometrical features.

On the other hand, a novel shape feature, fractal dimension \cite{reishofer2018age} is calculated by box-counting inside the tumor regions such as enhancing tumor and necrosis (see Figure \ref{fig:boxcount} (B) and (C)). It calculates the geometrical complexity of the brain tumor. Kurtosis, a non-gaussianity measurement of intensity probability distribution \cite{decarlo1997meaning}, is another essential feature. Entropy and histogram serve as the intensity features. Collectively, 71 features are extracted from the segmented tumor. Besides, we have age information and 1740 gene expression profiling data for each patient. 

\section{Experiments and Results}

\subsection{Dataset}
We use the TCGA GBM dataset to conduct our study. As BraTS-2017 \cite{menze2015multimodal, bakas2017advancing, bakas2017segmentation} consists of similar MRI dataset. Therefore we utilize it to train both our synthesis and segmentation models.

\subsubsection{BraTS}
We use BraTS-2017 \cite{menze2015multimodal, bakas2017advancing, bakas2017segmentation} to train our synthesis and segmentation model. It contains 285 and 46 cases in the training and validation set. Each case consists of 4 modalities, such as T1, T1c, Flair, and T2. There are 155 slices with a dimension of 240x240 in all the modalities. The MRI is annotated with three different regions of the tumor, such as necrotic and non-enhancing tumor (NCR/NET-label 1), edema (ED-label 2), and enhance tumor (ET-label 4).

\subsubsection{TCGA}
The Cancer Genome Atlas (TCGA) database consists of both MRI and genomic data. A previous study \cite{verhaak2010integrated} utilizes 202 patient cases with gene expression data, where 106 cases have corresponding MRI data. However, there is only one or two modalities of MRIs are available for most of the cases. Specifically, T2 modality is missing for the majority of the cases. There are also remarkable cases without T1 or Flair. We synthesize the missing modalities using MRI to MRI translation model (see section \ref{sec:mri_to_mri}). Moreover, the available MRIs are in different voxel sizes, spacing, and origin. We resample and register all the MRIs into equal voxel spacing similar to the BraTS-2017 dataset using ANTs library \cite{avants2011reproducible}. GBM gene expression data of TCGA is utilized to assay the glioblastoma tumor samples to get the ratio between the gene expression level to a known gene expression level  \cite{verhaak2010integrated}. The study
obtains 1740 genes with expression values after applying several filters. We exploit similar gene expression features to conduct all of our experiments.

\subsection{Implementation Details}
For both missing modality synthesis and Segmentation, the models have trained with the MRI slices of the BraTS-2017 training dataset. Learning rates are 0.0001 and 0.005 for the segmentation model and synthesis model (both generator and discriminator), respectively. The momentum and weight decay are 0.9 and 0.0001, respectively, for both and the adam algorithm is used for optimization. All the experiments are conducted with the Pytorch \cite{paszke2017automatic} deep learning framework. 

\subsection{Results}
\subsubsection{Missing Modalities Generation}

\begin{figure*}
\centering
\includegraphics[width=1\textwidth]{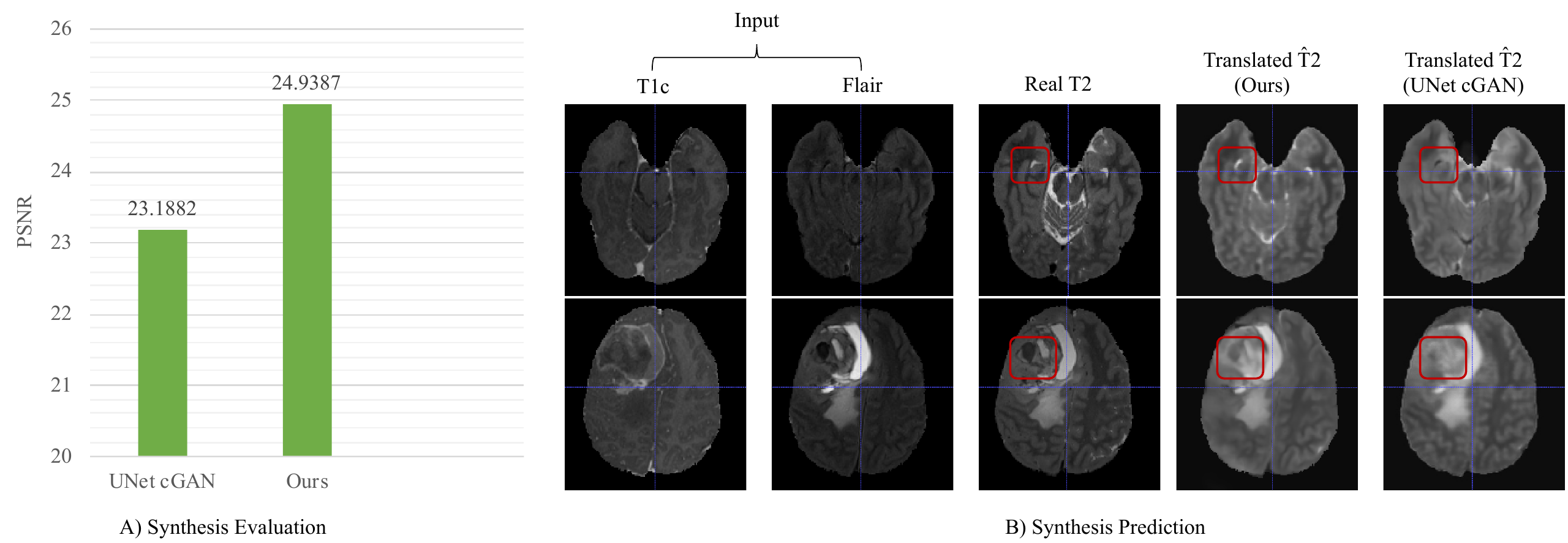}
\caption{Quantitative and qualitative results of cross modality generation results on BraTS-2017 MRI dataset. {(A)} PSNR value comparison between our model and UNet cGAN \cite{yang2018mri} to synthesize T2 from T1c and Flair. Validation is performed on the BraTS-2017 validation dataset. {(B)} Inputs and the images generated using our model and UNet cGAN\cite{isola2017image, yang2018mri}. Red boxes present poor synthesizing performance of UNet cGAN comparing to our model.}
\label{fig:PSNR}
\end{figure*}

We generate all the missing modalities of the TCGA dataset using cGAN with our FCN for the TCGA GBM dataset. The performance of the synthesis model is calculated using the peak signal to noise ratio (PSNR) \cite{hore2010image}. We also compare the accuracy of our model with original UNet cGAN \cite{isola2017image, yang2018mri}. The quantitative and qualitative results are shown in Figure \ref{fig:PSNR} (A) and (B). The red line shows the enhanced synthesizing performance of our model comparing to UNet cGAN \cite{isola2017image, yang2018mri}. We observe that the discriminator of encoder-decoder in our model performs better than just encoder discriminator in PatchGAN \cite{li2016precomputed}.

\subsubsection{Segmentation}

\begin{table}[!h]
\begin{center}
\caption{Mean Dice score \& Hausdorff value comparison on BraTS-2017 validation set. ET, WT, TC denote as  Enhancing Tumor, Whole Tumor, Tumor Core, respectively.}
\label{table:Seg_comp}
\begin{tabular}{|c|c|c|c|c|c|c|} 
%\begin{tabular*}{\tblwidth}{@{} LLLLLLL@{} }
\hline
\multirow{2}{*}{Model} &     \multicolumn{3}{c|}{Dice} &    \multicolumn{3}{c|}{Hausdorff} \\ \cline{2-7}  &  ET    & WT & TC & ET & WT & TC \\ \hline
Ours    &\textbf{0.7471}    & 0.8991    &\textbf{0.7991}  &\textbf{4.30}    &5.08   &6.78\\\hline
UNet scSE \cite{roy2018concurrent}        &0.7288    &0.8819     &0.7701           &5.71   & 6.37  &8.16 \\\hline
UNet \cite{ronneberger2015u}        &0.7243    &0.8776     &0.7615           &5.64    & 6.70  &8.23 \\\hline
EMMA  \cite{kamnitsas2017ensembles}    &{0.738} & \textbf{0.901}   &0.797    &{4.50} &\textbf{4.23}  &\textbf{6.56}\\ \hline
DeepLab v3 \cite{chen2018encoder} &    0.6631    & 0.8863 &    0.7834 & 4.39    & 5.62    & 8.51\\ \hline
GCN  \cite{peng2017large}   &0.7017    &0.8785    &0.7794       &5.22    &6.54   &8.37 \\ \hline
\end{tabular}
\end{center}
\end{table}

\begin{figure}
\centering
\includegraphics[width=0.48\textwidth]{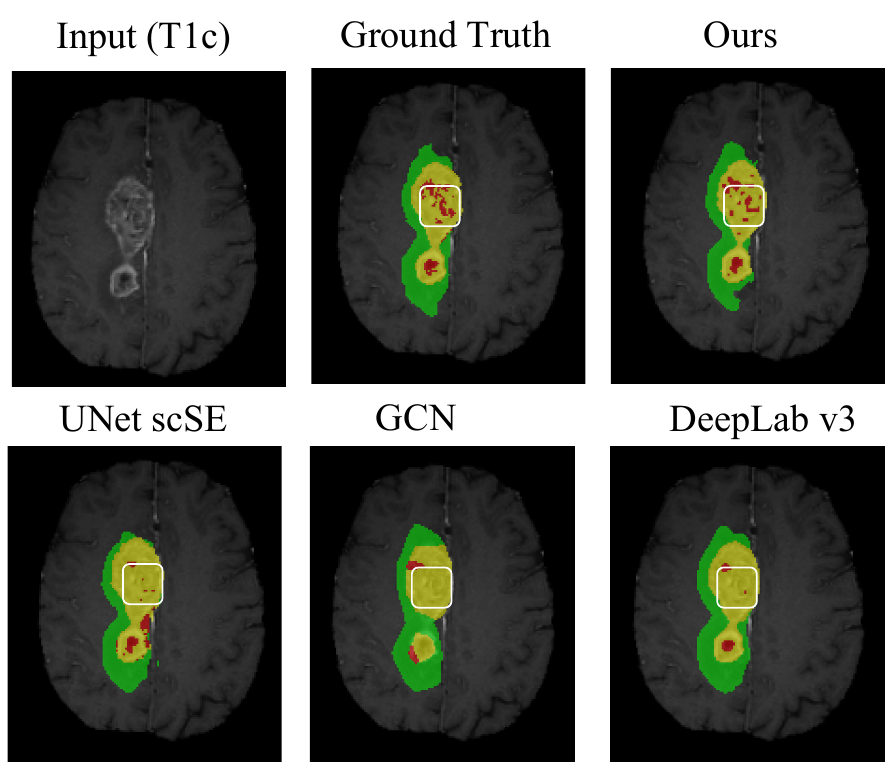}
\caption{Comparison between the ground truth and the obtained prediction from our model. A case is visualized randomly from cross-validation of BraTS-2017 training data. White boxes denote the improved prediction of necrotic parts by our model comparing
to other state-of-the-art segmentation models. Red, Yellow, and Green colors represent the tumor regions of Necrosis, Enhancing, Edema respectively.}
\label{fig:Segmentation_comp_brats}
\end{figure}

\begin{figure}
\centering
\includegraphics[width=.48\textwidth]{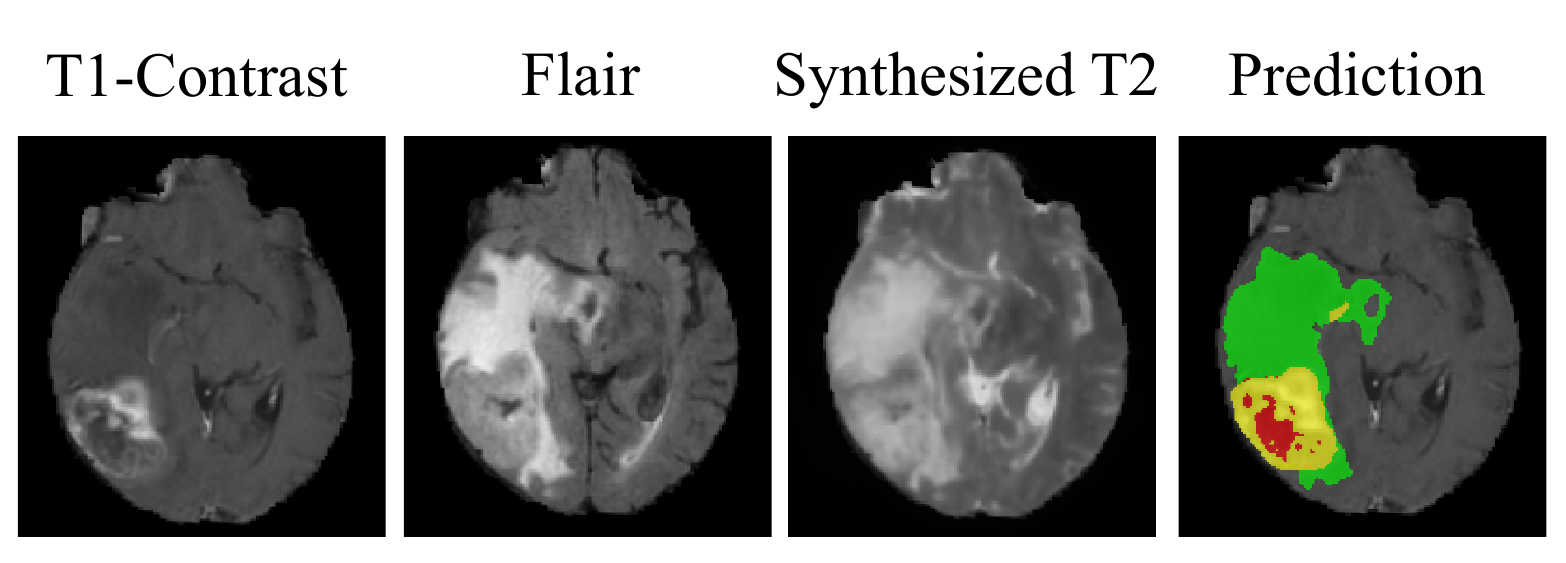}
\caption{ Visulization of the predicted segmentation for TCGA dataset using our model. The model is trained and validate with BraTS-2017 dataset and  synthesized and available modalities the TCGA data are exploited to predict the segmentation.}
\label{fig:Segmentation_results_tcga}
\end{figure}

We use TCGA available and synthesis MRI modalities of T1c, T2, and Flair to predict tumor regions from the trained model using the BraTS-2017 training set as 4-fold cross-validation. The prediction score of the validation set of BraTS-2017 is from the ensemble of all the models as \cite{isensee2017brain}.
There are three regions in the evaluation metrics, such as enhancing tumor(ET: class label 4), tumor core (TC: class labels 1,4), and whole tumor (WT: class labels 1,2,4). Table \ref{table:Seg_comp} shows the validation results of our model compared with several state-of-the-art models including the top rank model EMMA \cite{kamnitsas2017ensembles} in BraTS-2017 challenge. Our model achieves the competitive performance with EMMA and outperforms all the other models (e.g. Deeplab v3 \cite{chen2018encoder}, GCN \cite{peng2017large}, UNet \cite{ronneberger2015u}, UNet scSE
\cite{roy2018concurrent}), in both dice and Hausdorff scores. We observe that the tumor core is a crucial region to estimate survival. Qualitative visualization of the predicted segmentation of BraTS-2017 and TCGA dataset are shown in Figure \ref{fig:Segmentation_comp_brats} and \ref{fig:Segmentation_results_tcga}, respectively. 

\subsubsection{Survival Prediction}

\begin{figure*}
\centering
\includegraphics[width=1\textwidth]{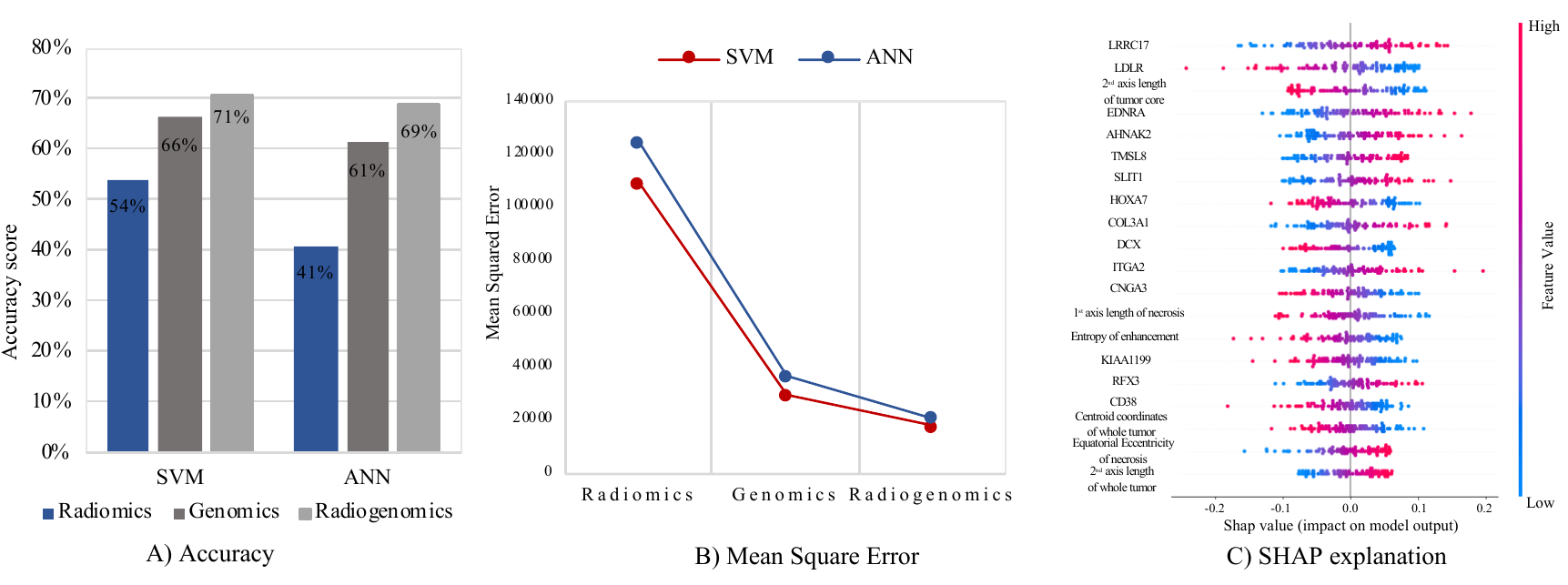}
\caption{Visualization of the performance comparison between ANN and SVM for radiomics, genomics, and radiogenomics. {(A)} Accuracy comparison between SVM and ANN. {(B)} Mean squared error for different features. ({C}) The impact of the most prominent features for the overall survival using SHAP value analysis \cite{lundberg2017unified}.}
\label{fig:OS_Shap}
\end{figure*}

Our radiogenomic model of SVM and ANN are evaluated with regression and classification metrics using RFE selected features of 8 radiomics, 43 genomics, and fused 51 radiogenomics. Table \ref{table:ANN_preformance} shows the mean square error (MSE) of the survival prediction days and accuracy, sensitivity, and specificity of the classification of short, medium, and long survival prediction. The maximum accuracy achieves with 84.62\% on model-3 among 4-fold cross-validation. Figure \ref{fig:OS_Shap} (A) and (B) represent the comparative performances of the ANN and SVM models and the impact of genomic features in OS estimation. The best performing ANN model consists of 6 hidden nodes (a single layer) with a ReLU activation function. As a state-of-art local explanation method, we use the Shapley Additive explanations (SHAP) \cite{lundberg2017unified}, highlighting the essential features of the best performing model. Figure \ref{fig:OS_Shap} (C) represents the mean absolute value of the SHAP values for each prominent feature according to the contribution to the model performance and each class. Our results show that the performance boosts to almost double after fusing genomic features with radiomic. We observe that the ANN model often overfit and performs poor comparing to SVM.

\begin{table*}[!h]
%Table \ref{table:os_ANN_c}
\begin{center}
\caption{4-fold cross validation results of the radiogenomic features using SVM. Short, medium and long survival rate are abbreviated to S., M., and L. respectively. MSE, and ACC represent mean square error, and accuracy respectively.}
\label{table:ANN_preformance}
\begin{tabular}{|c|c|c|c|c|c|c|c|c|}
%\begin{tabular*}{\tblwidth}{@{} LLLLLLLLL@{} }
\hline
\multirow{2}{*}{fold} & \multirow{2}{*}{MSE} &\multirow{2}{*}{Acc(\%)} & \multicolumn{3}{c|}{Sensitivity (\%)}  & \multicolumn{3}{c}{Specificity (\%)}\\ %\cline{4-9} 
\cline{4-9}
& & & S & M & L &S & M & L \\ 
\hline
1&    25001&    50.00 & 40    &16.67&    80&    93.75&    60    &75 \\ \hline
2&    22384&    76.92 &60    &66.67&    100 &    87.50&    80    &100 \\ \hline
3&    \textbf{11731}&    \textbf{84.62} & 90    &66.67    &90&    93.75&    90    &93.75 \\ \hline
4&    13982&    73.08 &90    &16.67&    90    &87.50&    90&    81.25 \\ \hline

Avg & {18275}    & {71.15} &70&    41.67&    90    &90.63&    80&    87.50 \\ \hline
\end{tabular}
\end{center}
\end{table*}

\section{Discussion}
Our model achieves a significant margin in generating missing modalities and segmenting the GBM regions from the experimental results. However, the study in \cite{lee2019contrast} demonstrates that there are remarkable challenges of GAN based image translation models to learn contrast MRI (T1c) due to a lack of scalability. We also observe poor performance in synthesizing T1c from Flair. In these cases, we only exploit available modalities to predict the tumor segmentation. Further, we validate the performance improvements in segmentation by adding one by one modality, as in Table \ref{table:modality}.

To estimate the survival duration, we train regression models such as SVM, ANN with the most prominent geometric and shape features of the tumor, and combine with gene expression profiling data. We observe that gene expression profiling has a high correlation with the GBM progression and fused radiogenomic features boosts the prediction accuracy. Figure \ref{fig:OS_Shap} (C) y-axis represents the 20 most significant features of our radiogenomic model. Genomic features such as LRRC17 and LDLR demonstrate the highest impact where radiomic features such as axis coordinates of WT and TC and histogram have a remarkable influence in OS prediction. Moreover, overexpression of LRRC17, EDNRA, AHNAK2, TMSL8, and SLIT1 genes have a positive impact on the model output, leading to a more prolonged overall survival of GBM patients. High expression of the LDNRA gene impacts the model by reducing the overall survival of GBM patients. Our work verifies that fusing gene and radiomic biomarkers are significant for an accurate survival prediction of glioma patients, consistent with the previously done studies \cite{chaddad2018novel}.
One of the limitations of this study is the limited dataset since we perform regression in the overall survival prediction task. Therefore image synthesis to generate the missing MRI modalities is important to increase the population cohort in this study, further enhancing the model performance.   

\begin{table}[!h]
%Table \ref{table:modality}
\begin{center}
\caption{Dice score comparison for different number of input modalities for the UNet \cite{ronneberger2015u} segmentation model.}
\label{table:modality}
\begin{tabular}{|c|c|c|c|}
%\begin{tabular*}{\tblwidth}{@{} LLLL@{} }
\hline
\multirow{2}{*}{Number of modalities} & \multicolumn{3}{c|}{Dice} \\ \cline{2-4} 
& ET    & WT & TC  \\ \hline
1 modality (T1C) &0.6557 &    0.6829 &0.6872\\ \hline
2 modalities (T1C, Flair)&    0.7132&    0.8758&    0.7516 \\ \hline
3 modalities(T1C, Flair, T2)&    \textbf{0.7243}&   
\textbf{0.8776}&    \textbf{0.7615} \\ \hline
\end{tabular}
\end{center}
\end{table}

\section{Conclusion}
In this work, we propose a novel FCN model for MRI missing modality synthesis, and tumor segmentation, and radiogenomic models for the overall survival estimation of GBM patients. Our FCN forms with octave convolution and skip-scSE module improve the prediction accuracy by eliminating feature redundancy and sparsity in model learning. The proposed FCN uses as a generator in a cGAN with a modified discriminator to synthesize the missing modalities. The quantitative and qualitative results indicate that our model outperforms other state-of-the-art models of synthesis and segmentation in most of the evaluation metrics. Our research demonstrates that radiogenomics is a promising strategy for OS prediction for the noninvasive management of GBM. Further, gene mutation data and other gene biomarkers can integrate with radiomic features to estimate OS in future work.

\bibliographystyle{IEEEtran}
\bibliography{mybib}

\end{document}